\documentclass[fleqn,12pt,twoside]{article}
\usepackage{espcrc1}

\usepackage{graphicx}
\usepackage[figuresright]{rotating}

\newcommand{\AmS}{{\protect\the\textfont2
  A\kern-.1667em\lower.5ex\hbox{M}\kern-.125emS}}

\title{Observations of Neutron-Capture Elements in the Early Galaxy}

\author{
C. Sneden\address[TEXAS]{Department of Astronomy and McDonald
Observatory, University of Texas, Austin, TX 78712, USA}
\thanks{ The authors all thank our colleagues who have contributed 
to the results presented here.  Research supported by U.S. NSF 
grant AST-9987162.},
J. J. Cowan\address[OKLA]{Department of Physics and Astronomy, 
University of Oklahoma, Norman, OK 73019, USA},\addressmark[TEXAS]
\thanks{Research supported by U.S. NSF grant AST-9986974.},
and J. E. Lawler\address[WISC]{Department of Physics,
University of Wisconsin, Madison, WI 53706 USA}
\thanks{Research supported by U.S. NSF grant AST-981940.}
}

\begin{document}

\maketitle

\begin{abstract}
Neutron-capture elements in low metallicity Galactic halo stars 
vary widely both in overall contents and detailed abundance patterns.
This review discusses recent observational results on the n-capture elements,
discussing the implications for early Galactic nucleosynthesis of:
(a) the star-to-star ``bulk'' variations in the n-capture/Fe abundance
ratios; (b) the distinct signature of rapid n-capture synthesis
events in many (most?) of the lowest metallicity stars; (c) the existence
of metal-poor stars heavily enriched in the products of slow n-capture
synthesis reactions; and (d) the now-routine detection of radioactive
thorium (and even uranium in one and possibly two cases) in the spectra 
of metal-poor stars.
\end{abstract}

\section{NEUTRON-CAPTURE ELEMENTS}

Neutron-capture elements are those heavy elements whose isotopes are 
primarily synthesized in neutron bombardment reactions; these include 
all elements with Z~$>$~30.
The relative rates of neutron ingestion by nuclei compete with $\beta$-decay
rates to determine the final abundance distribution.
If the neutron flux is small enough that all unstable nuclei may undergo
$\beta$-decay between successive neutron captures, element synthesis proceeds
up the valley of $\beta$ stability. 
This slow neutron addition is called the $s$-process, and occurs principally
in the He-burning zones of low and intermediate mass AGB stars.
If however the neutron flux is very large compared to $\beta$-decay rates,
extremely neutron-rich nuclei are created which then rapidly undergo
$\beta$-decay back toward stability.
This rapid neutron-capture ($n$-capture) synthesis, called the $r$-process, 
can occur in several situations during supernova deaths of high-mass 
stars and their creation of neutron-star remnants.

Isotopes of the $n$-capture elements may be created by the 
$s$-process, the $r$-process, or both.
The solar-system abundances of these isotopes have been built in 
roughly equal measure of products of the $r$- and $s$-process.
The manner in which isotopic (and hence elemental) abundance fractions 
are usually determined \cite{cam82,kap89,bur00} deserves comment here.
In the $s$-process the product of isotopic abundance and neutron-capture
cross-section is a well-defined slowly varying function of atomic mass,
which can be matched to the observed (meteoritic) abundances of isotopes
that only can be made in the $s$-process. 
The $s$-process abundances of other isotopes can then be predicted from
this match.
Then the $r$-process abundances are either taken directly from
meteoritic data for $r$-only isotopes, or computed by subtraction of
the $s$-process amounts from the total abundances for those isotopes
that can be synthesized in both $n$-capture processes.
Finally, since only total elemental abundances may normally be determined
from stellar spectra, the isotopic $r$- and $s$-process isotopic
abundances for each element are summed.

\begin{figure}[htb]
\includegraphics[width=35pc]{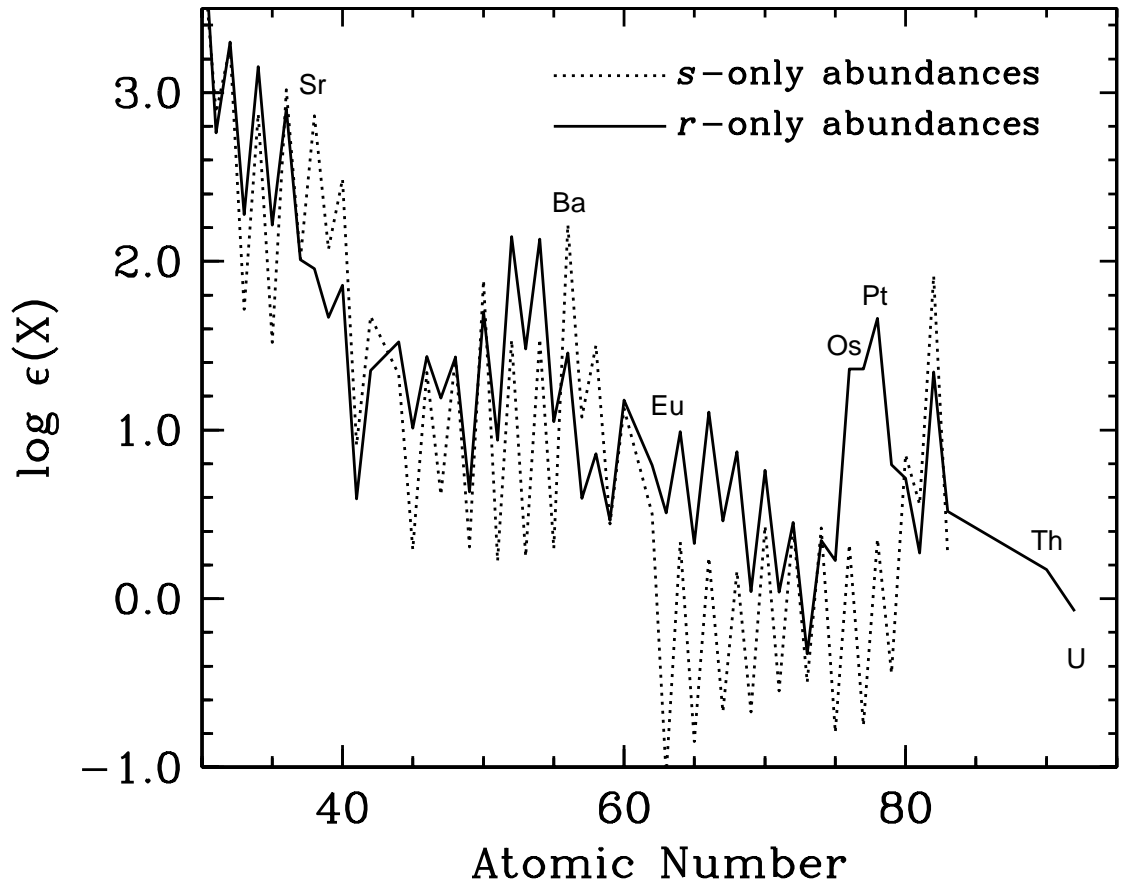}
\caption{Abundance distributions of $n$-capture elements in the solar
system, separated into their $r$-process-only and $s$-process-only
components \cite{bur00}, as functions of atomic number Z.
Abundances are given in standard stellar spectroscopic notation: 
log~$\epsilon$(A)~$\equiv$ log$_{\rm 10}$(N$_{\rm A}$/N$_{\rm H}$)~+~12.0.}
\label{rands}
\end{figure}

A recent re-assessment \cite{bur00} of the solar-system elemental 
abundances is shown in Figure~\ref{rands}.
With this figure a caution must be given.
For those elements whose solar-system abundances are dominated by the
$s$-process (e.g. Sr, Ba, La), the small $r$-process component obtained
in the subtraction process carries with it the abundance uncertainty
estimate of the $s$-process component, and this value can in extreme
cases be as large as the inferred $r$-process abundance.
Thus the solar-system $r$-process abundance set should be treated with
some caution.
But for the foreseeable future this distribution will remain the
standard, because direct predictions of $r$-process abundances (e.g.,
\cite{gor97,cow99}) must rely on mass models of extremely neutron-rich 
nuclei, almost none of which has ever been experimentally detected.
Theoretical assessments of the solar-system $s$- and $r$-process
abundances have also been made (e.g. \cite{arl99}), and these yield
similar results to those displayed in Figure~\ref{rands}).

\section{NEUTRON-CAPTURE ELEMENTS AT LOW METALLICITY} 

A dominant aspect of $n$-capture element abundances in very metal-poor 
halo stars is their extremely large variability ``in bulk'' with respect
to the Fe-peak elements (for early indications compare the abundances
reported for HD~122563 \cite{wal63} with those of HD~115444 \cite{gri82},
and for larger-sample surveys see, e.g., \cite{gil88,mcw95,rya96,car02}).
The observed star-to-star range in [$<n$-capture$>$/Fe] grows with decreasing
metallicity, exceeding observational errors at [Fe/H]~$\approx$ --2.0
and reaching a scatter factor of more than a thousand below
[Fe/H]~$\approx$ --2.5 (e.g., Figure~12 of \cite{car02}).
The lowest metallicity stars are usually faint and high resolution
spectra of them requires the use of 8m-class telescopes.  

\begin{figure}[htb]
\includegraphics[width=33pc]{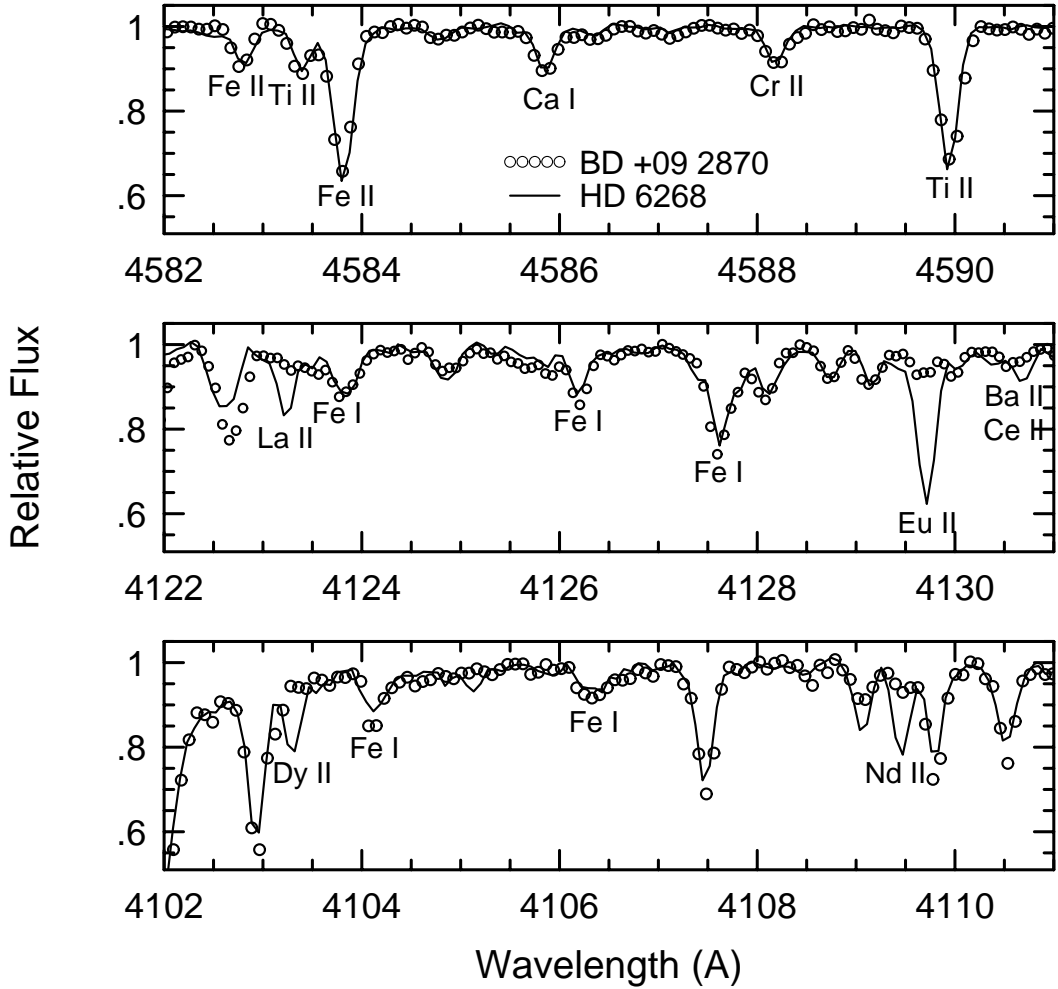}
\caption{Comparison of the spectra of two metal-poor giant stars
in a wavelength region dominated by Fe-peak species features (top panel),
and in two regions with significant $n$-capture features.
This figure is adapted from Figure~3 of \cite{bur00}.
These two stars have very similar atmospheric parameters and
metallicities: T$_{eff}$~= 4650~K, log~g~= 1.50, v$_t$~= 1.7 km~s$^{-1}$,
[Fe/H]~= --2.4.}
\label{threespec4}
\end{figure}

Proof of the star-to-star $n$-capture scatter can be had
through simple spectrum comparisons of relatively bright stars,
as has been done by \cite{bur00}, whose Figure~3 is adapted as our
Figure~\ref{threespec4}.
The two stars shown here have almost identical atmospheric parameters, 
and the spectral region of the top figure panel suggests
that their metallicities and light element abundance ratios are about
the same.
But the $n$-capture element features have strengths that differ
by nearly a factor of ten, and it is easy to show that derived abundances
of these elements differ by the same amount with little dependence
on atmospheric modeling uncertainties.
And the large star-to-star scatter can probably have only one interpretation:
the influence of individual and localized nucleosynthesis events in
short-lived stars in a poorly-mixed early Galactic halo.

A second feature of $n$-capture elements in metal-poor stars is the
general trend toward pure $r$-process signatures (at least for
elements with Z~$\geq$~56) at lowest metallicities.
Significant exceptions to this trend are known to exist. 
In particular
there are some stars with [Fe/H]~$<$ --2.0 with large overabundances
of C and $s$-process-dominated elements such as Sr, Ba and even Pb
(e.g. \cite{eck01,aok02}).
Such objects may have received these enhanced abundances via mass transfer
from a former AGB companion star.
A more thorough discussion of these important stars has been given by
S. G. Ryan at this conference, and will not be commented upon further here.

For more ``normal'' low metallicity stars, a downward trend in [Ba/Eu]
is observed with decreasing metallicity (the pioneering observational
and theoretical work here were done by \cite{spi78,tru81}, and for
a recent summary see Figure~3 of \cite{cow01} and Figure~7 of \cite{bur00}). 
This decrease, which tends toward [Ba/Eu]~$\approx$ --1.0 below 
[Fe/H]~$\approx$ --2.5, is qualitatively and quantitatively in agreement
with essentially complete absence of products of $s$-process synthesis
in the most metal-poor stars.  
The trend combined with the large star-to-star scatter in the bulk amounts
of the $n$-capture elements noted above rounds out the suggested picture
of $r$-process synthesis occurring in and around the supernova deaths
of high mass, very low metallicity stars, and seeding their local ISM
out of which quickly formed the metal-poor stars that we observe today.

Unfortunately another undeniable aspect of the [Ba/Eu] trend with metallicity
is its large scatter at any [Fe/H] value.  
It is possible (even likely) that some of the scatter is intrinsic
to the stars, caused by $s$-processing even in the early Galaxy.
Some of the very metal-poor stars with [Ba/Eu] ratios much greater than the
$r$-process value probably have undergone some a transfer episode from
a higher mass companion which is now a compact object.
But a lot of the [Ba/Eu] scatter must be due to observational and 
analytical uncertainties, particularly in the derivation of Ba abundances.
Usually just the very few strong Ba~II low-excitation lines are available
for analysis, and these features are saturated in cool giants that are 
most often studied at low metallicities.

There are two approaches that can mitigate this problem.
First, Ba abundances should be determined for main sequence stars,
as has been recently done in \cite{car02}.  
The Ba~II lines are intrinsically much weaker in upper main sequence
stars than they are in giants of similar metallicity. 
The problem with this approach lies in simultaneous detection of
Eu~II features, which become almost undetectably weak in very low metallicity
main sequence stars.
Second, instead of using Ba~II lines, a switch could profitably be made to
analyses of La~II (which is also predominantly an $s$-process product
in solar-system material).
This species offers several analytical advantages: (a) La has just one 
stable isotope, compared with five abundant ones for Ba; (b) La~II lines are
more numerous and weaker than Ba~II lines, and usually are of comparable
strengths to those of Eu~II; and (c) accurate transition probabilities and
hyperfine structure parameters are known for many La~II lines \cite{law01a}.
Preliminary results of a new large-sample study of [La/Eu] have been
given in \cite{sim02}, and they show a smaller star-to-star scatter
than does [Ba/Eu] and the same trend to a pure $r$-process ratio
at lowest metallicities.
The use of La~II instead of Ba ~II lines is recommended for judging the
$s$-process contents of all stars where the $n$-capture features all
are strong.

\section{DETAILED ABUNDANCE DISTRIBUTIONS IN $r$-PROCESS-RICH STARS}

A few very metal-poor stars have been discovered to have both very large
overall $n$-capture element contents (here arbitrarily defined as
[Eu/Fe]~$\sim$ +1 or greater) and nearly pure $r$-process signatures 
among these elements.
Such stars include HD~115444 ([Fe/H]~= --2.9, [Eu/Fe]~= +0.9 \cite{wes00}),
CS~22892-052 ([Fe/H]~= --3.1, [Eu/Fe]~= +1.6 \cite{sne00}),
BD+17~3248 ([Fe/H]~= --2.1, [Eu/Fe]~= +0.9 \cite{cow02}),
and CS~31082-001 ([Fe/H]~= --2.9, [Eu/Fe]~= +1.6 \cite{hil02}).
Such stars are excellent laboratories for the study of $n$-capture
elements because their normally dominant Fe-peak transitions are
very weak in these low metallicity stars, and the relative overabundances
of the $n$-capture elements reveals many spectral features that
are normally hidden in stellar spectra.
For example, Tb~II lines are much more readily identified in the
spectrum of CS~22892-052 than they are in the solar spectrum \cite{law01b}.

\begin{figure}[htb]
\includegraphics[width=35pc]{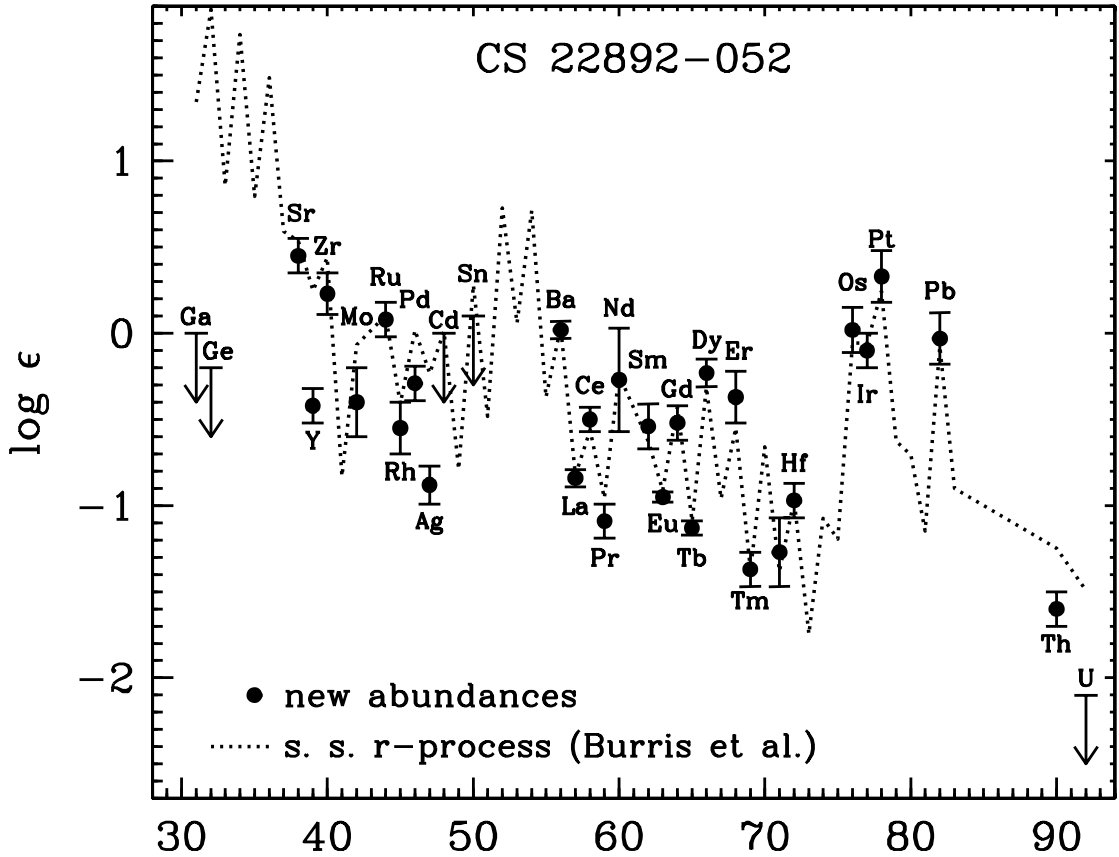}
\caption{Observed $n$-capture abundances \cite{sne02} in the halo giant star
CS~22892-052 compared to the $r$-process solar-system abundance distribution
of Figure~1, scaled to match the observed Eu abundance.
The error bars on the observed points represent the sample standard
deviations of the abundances of each element.}
\label{abnic7}
\end{figure}

In Figure~\ref{abnic7} we summarize the current \cite{sne02} CS~22892-052
$n$-capture elemental abundance set, comparing it
to the scaled solar-system abundance distribution \cite{bur00}.
The abundance matches shown here lead to the following brief comments,
reemphasizing some conclusions from our previous contributions.

First, for the 17 observed stable elements with Z~$\geq$~56, a near-perfect 
match exists with the scaled solar-system $r$-process curve.
The fit is good enough that elements which either deviate
from the solar curve (e.g. Er) or have large line-to-line scatter (e.g. Nd)
point to areas for further study in nuclear and atomic physics,
respectively.
Comparing the abundances shown here to those of the first large abundance
study of CS 22892-052 \cite{sne96}, the major sharpening of the
newer abundance results (smaller line-to-line scatters) may be traced to
several recent extensive lab studies of rare-earth elements (for 
reviews and comments on recent work in this area see \cite{wah02,sne02b}).
Note finally that the excellent agreement with scaled solar system
abundances now has been extended to isotopic abundances for Eu
in three $r$-process-rich stars \cite{sne02c}.

Second, elements of the third $n$-capture peak (76~$\leq$~Z~$\leq$~82)
have their transitions mainly below 3500~\AA, and thus some of these
abundances have been obtained from HST STIS observations.
It is more difficult to derive reliable abundances in the UV because
even very metal-poor giants have rich atomic and molecular line spectra
in this wavelength regime.
Even so, the abundances of this element group are consistent with
the same scaled solar curve that matches the abundances of the rare-earth
elements.

Third, the long-lived radioactive element Th is detected in CS~22892-052 
and indeed it has been observed in many metal-poor giants \cite{fra93,joh01}.
The derived Th abundance, when ratioed to the abundance of a stable
$n$-capture element (usually taken to be Eu) can lead to age estimates
for the oldest stars of the Galaxy.
The prospects and problems of such nucleocosmochronology have been
discussed in several papers (e.g. \cite{cow99,gor01,sch02}) and space
does not permit further discussion here.
An exciting new development has been the detection of U, a second 
chronometer element, in CS~31082-001 \cite{hil02} and its ratio with 
respect to Th indicates a decay age of about 15~Gyr, in agreement
with other kinds of Galactic age estimates.
See also \cite{cow02} for an additional tentative detection of U in
a metal-poor halo star and a similar radioactive age estimate.

Fourth, lighter $n$-capture elements, those with 38~$\leq$~Z~$\leq$~50, 
show some deviations from the scaled solar $r$-process distribution, 
particularly for Y, Mo, Pd, and Ag.
One caution must be given here, that for most of these elements the number
of lines participating in the derived abundance is significantly smaller
than the number available for abundance determinations of rare earth
elements; hence the generally larger error estimates that we attach
to these points.
Suggestions that the $r$-process nucleosynthetic history of the lighter
$n$-capture elements may be complex began with \cite{was96} and has
no clear resolution today.
The deviations may signal the presence of a second type of $r$-process as
proposed in \cite{qia00}.
Alternatively, models have been suggested that can synthesize the whole
range of these observed abundances in one site (e.g. \cite{cam01})

Fifth, the lightest elements attributed mostly to $n$-capture processes
cannot be detected on our CS~22892-052 spectra, and the derived upper
limits to their abundances are more than an order of magnitude below
the scaled solar $r$-process curve.
Clearly these two elements either owe their origin to a different 
nucleosynthetic process or the $r$-process simply may have pushed beyond
these elements.

\section{SUMMARY}

A rich variety of $n$-capture abundance distributions has been found
in low metallicity Galactic halo stars, and discoveries of new stars
with unique chemical composition mixes show no signs of abating.
Early Galactic nucleosynthesis has littered the halo with many clues,
and it will take the combined efforts of astronomers working with halo
low resolution surveys, stellar spectroscopists, lab and theoretical 
atomic physicists, experimental and theoretical nuclear physicists,
and supernova modelers to properly decode the messages contained in
the $n$-capture abundances of metal-poor stars.

\end{document}